\documentclass[11pt,a4paper]{article}
\usepackage{jheppub}
\usepackage{amsmath,amssymb,theorem,epsfig,url,psfrag,eepic,mathtools,mathrsfs,amsbsy,dsfont}
\usepackage{indentfirst}
\usepackage{graphicx}
\usepackage{tikz,pgfplots}
\pgfplotsset{compat=1.17}
\begin{document}
\title{\textbf{On $\beta$-function of $N=2$ supersymmetric integrable sigma-models}\vspace*{.3cm}}
\date{}
\author[a,b]{Mikhail Alfimov,}
\author[c]{Ivan Kalinichenko}
\author[d,e]{and Alexey Litvinov}
\affiliation[a]{HSE University, 6 Usacheva str., Moscow 119048, Russia}
\affiliation[b]{P.N. Lebedev Physical Institute of the Russian Academy of Sciences, 53 Leninskiy pr., Moscow 119991, Russia}
\affiliation[c]{
Institute for Theoretical and Mathematical Physics, Moscow State University, 119991, Leninskie Gory, GSP-1, Moscow, Russia}
\affiliation[d]{Skolkovo Institute of Science and Technology, Krichever Center for Advanced Studies, Bolshoy Boulevard 30, bld. 1, Moscow 143026, Russia}
\affiliation[e]{Landau Institute for Theoretical Physics, 1A Akademika Semenova av.,  Chernogolovka 142432, Russia}
\emailAdd{malfimov@hse.ru}
\emailAdd{kalinichenko.ia18@physics.msu.ru}
\emailAdd{litvinov@itp.ac.ru}
\abstract{
We study regularization scheme dependence of K\"ahler ($N=2$) supersymmetric sigma models. At the one-loop order the metric $\beta$ function is the same as in non-supersymmetric case and coincides with the Ricci tensor. First correction in MS scheme is known to appear in the fourth loop \cite{Grisaru:1986px,Grisaru:1986dk}. We show that for certain integrable K\"ahler backgrounds, such as complete $T-$dual of $\eta$-deformed $\mathbb{CP}(n)$ sigma models, there is a scheme in which the fourth loop contribution  vanishes.
}
\maketitle
%%%%%%%%%%%%%%%%%%%%%%%%%%%%%%%%%%%%%%%%%%%%%%%%%%%%%%%%%%%%%%%%%%%%%%%%%%%%%%%%%%%%%%%%%%%%%%%
%%%%%%%%%%%%%%%%%%%%%%%%%%%%%%%%%%%%%%%%%%%%%%%%%%%%%%%%%%%%%%%%%%%%%%%%%%%%%%%%%%%%%%%%%%%%%%%
%%%%%%%%%%%%%%%%%%%%%%%%%%%%%%%%%%%%%%%%%%%%%%%%%%%%%%%%%%%%%%%%%%%%%%%%%%%%%%%%%%%%%%%%%%%%%%%
\section{Introduction}
Two-dimensional sigma models 
\begin{equation}
  S[\boldsymbol{X}]=\frac{1}{4\pi}\int \left(G_{ij}(\boldsymbol{X})\partial_{a}X^{i}\partial_{a}X^{j}+\dots\right)\,d^{2}\boldsymbol{x}\,.
\end{equation}
are renormalizable in the weak sense, meaning  that RG flow engage in general an infinite number of coupling constants. Staying withing the class of pure metric bosonic sigma-models the RG flow equation takes the form  \cite{Ecker:1971xko,Friedan:1980jm}
\begin{equation}\label{RG-equation}
    \dot{G}_{ij}+\nabla_iV_j+\nabla_jV_i=-\beta_{ij},\qquad
    \beta_{ij}=R_{ij}+O(R^2).
\end{equation}
With the higher loop terms dropped (of order $O(R^2)$) this equation is nothing else, but the celebrated Ricci flow equation. From the analysis of this non-linear equation one obtains that typically starting from any smooth initial metric the solution  develops singularities both forward  (IR)  and backward (UV) in the RG time. However, it is well known that this equation also possesses ancient solutions that exist on semi-interval starting from $t=-\infty$. Such solutions correspond to asymptotically free quantum field theories and thus are of primary interest. 

The most well known examples of ancient solutions correspond to sigma-models on compact symmetric spaces $G/H$ which are integrable on both classical and quantum level (provided that $H$ is simple). Recent studies of integrable deformations of $2D$ sigma-models (see \cite{Hoare:2021dix} for pedagogical review to the subject) provide new examples of ancient solutions. These new solutions are expected to correspond to integrable quantum field theories. In the case of the so called $\eta-$deformed $O(N)$ sigma-models this has been conjectured and checked in \cite{Fateev:2018yos,Litvinov:2018bou} (see also \cite{Fateev:2019xuq,Litvinov:2019rlv,Alfimov:2020jpy} for other models). One of the main outcomes of \cite{Fateev:2018yos,Litvinov:2018bou,Fateev:2019xuq,Litvinov:2019rlv,Alfimov:2020jpy} is that $\eta-$deformed theory is integrable QFT whose factorized $S$-matrix coincides with trigonometric version of Zamolodchikov's $S-$matrix for $O(N)$ theory \cite{Zamolodchikov:1978xm}.

The approach advocated in \cite{Fateev:2018yos,Litvinov:2018bou,Fateev:2019xuq,Litvinov:2019rlv,Alfimov:2020jpy} is somewhat indirect. It uses as an assumption that the classical action of $\eta-$deformed sigma-model gets promoted to the full quantum action, when all higher loop orders of the RG equation are taken into account. In particular it implies that there should exist ancient solution to the full RG equation \eqref{RG-equation}\footnote{In fact its version with the non-trivial $B$-field contributions which are present in the case of $\eta-$deformed sigma-models.}. The question of the existence of such solutions is far from being trivial. One of the main obstacles lies in the fact that the right hand side of \eqref{RG-equation} -- the $\beta$-function is not known in general form. Saying it more accurately, what is not known is the normal form of the $\beta-$function to which it can be reduced by the choice of regularization scheme. This is to be compared  to the situation happening  for the theories with finite number of couplings where, say for the theory with one classically marginal coupling, the normal form of the $\beta$-function is given by first two loop orders.

The question of higher loop corrections to integrable sigma-model backgrounds has been considered first in \cite{Hoare:2019ark,Hoare:2019mcc,Hoare:2020fye}. It has been noticed that classical background require quantum corrections in order to satisfy  \eqref{RG-equation} in higher loops. In \cite{Alfimov:2021sir} two of the present authors have studied regularization scheme dependence of the $\beta$-function for sigma models with two-dimensional target space. It has been shown that at four-loop approximation one can pick a scheme in which the $\beta$-function for bosonic sigma-model has the form\footnote{We note that in the case of two-dimensional target space there are simplifications due to the relation
\begin{equation}\label{Riemann-2D}
    R_{ijkl}=\frac{R}{2}\left(G_{ik}G_{jl}-G_{il}G_{jk}\right)\,,
\end{equation}
which allows to express all terms in the $\beta$-function through the scalar curvature and the metric.}
\begin{multline}\label{all-loop-beta-function-first-4-loops}
\beta_{ij}=\left(\frac{R}{2}+\frac{R^2}{4}+\frac{3R^3}{16}+\frac{5R^4}{32}-\frac{2+\zeta(3)}{64}\nabla^2\left(R^3+2R\nabla^2R-\frac{1}{2}\nabla^2R^2\right)+\dots\right)G_{ij}-\\-
\left(\frac{1}{16}+\frac{5R}{32}+\dots\right)\nabla_{i}R\nabla_{j}R+\ldots\,.
\end{multline}
It is important that the $\zeta(3)$ terms in the $\beta-$function can not be exterminated by the choice of scheme. However, as it was shown in \cite{Alfimov:2021sir}, for integrable sigma-models such as $\eta-$deformed $SU(2)/U(1)$ model (also known as sausage model \cite{Fateev:1992tk}), and the corresponding $\lambda$-deformed model \cite{Sfetsos:2013wia,Hollowood:2014rla}, this term vanishes due to the special relation valid for the classical action
\begin{equation}\label{R3-relation}
    R^3+2R\nabla^2R-\frac{1}{2}\nabla^2R^2=\hbar^3\big(\kappa-\kappa^{-1}\big)^2\big(\kappa+\kappa^{-1}\big),
\end{equation}
where $\kappa$ and $\hbar$ are parameters of the classical action. Using this fact, the authors of \cite{Alfimov:2021sir} proposed "exact" expressions for $\eta$ and $\lambda$ deformed backgrounds which solve \eqref{RG-equation} in the four-loop approximation.

This paper is aimed to generalize the results of \cite{Alfimov:2021sir} and in particular the relation \eqref{R3-relation} to higher dimensional target spaces. We shall consider the case of $N=2$ supersymmetric pure metric sigma-models. This choice is dictated by simplification reasons. First of all, the $\beta$-function of supersymmetric model (even for $N=1$ model) does not have two and three loop contributions in the MS scheme \cite{Grisaru:1986px,Grisaru:1986dk}
\begin{equation}\label{SUSY-4-loop-intro}
\beta_{ij}^{\scriptscriptstyle{\textrm{SUSY}}}=R_{ij}+T_{ij}^{\scriptscriptstyle{\textrm{$4$-loop}}}+\dots
\end{equation}
Moreover $N=2$ supersymmetry requires the target space manifold to be K\"ahler. In this case the number of scheme parameters reduces drastically and makes the analysis relatively simple. We will show that there exist a special scheme (see eqs. \eqref{new-K} and \eqref{delta-K-tilde} below) in which the fourth-loop contribution vanishes for a class of integrable sigma-models known as complete $T-$dual of $\eta-$deformed $SU(n)/U(n-1)$ sigma-models. 

It is important to note that the supersymmetric version of the complete $T-$dual of $\eta-$deformed $SU(n)/U(n-1)$ sigma-model, contrary to its bosonic version, is  integrable at the quantum level. This model corresponds to integrable perturbation of $N=2$ Kazama-Suzuki coset CFT \cite{Kazama:1988qp}.  Moreover, one can make arguments justifying  that there should exist a scheme in which this model is one-loop exact (see section \ref{Tdual-CPn}).

This paper is organized as follows. In section \ref{beta-sec} we discuss scheme dependence of the $\beta$-function in $N=2$ sigma-models. We will emphasize the importance of using only those metric redefinitions that correspond to the so called K\"ahler tensors \cite{AlvarezGaume:1980dk}. We also provide an explicit scheme, which we call integrable scheme,  that does not alter two and three loop contributions, but changes the fourth loop one.  In section \ref{Tdual-CPn} we show that in integrable scheme the fourth-loop contribution vanishes for a class of integrable K\"ahler sigma-models known as complete $T-$dual of $\eta-$deformed (Yang-Baxter deformed) $SU(n)/U(n-1)$ sigma-model. We also provide some arguments in favor of one-loop exactness of the supersymmetric version of this model. In section \ref{conclusion} we give the concluding remarks as well as list some open questions.
%%%%%%%%%%%%%%%%%%%%%%%%%%%%%%%%%%%%%%%%%%%%%%%%%%%%%%%%%%%%%%%%%%%%%%%%%%%%%%%%%%%%%%%%%%%%%%%%%%%%%%%%%%%%%%%%%%%%%%%%%%%%%%%%%%%%%%%%%%%%%%%%%%%%%%%%%%%%%%%%%%%%%%%%%%%%%%%%%%%%%%%%%%%%%%%%%%%%%%%%%%%%%%%%%%%%%%%%%%%%%%%%%%%%%%%%%%%%%%%%%%%%%%%%%%%%%%%%%%%%%%%%%%%%%%
\section{Beta-function for \texorpdfstring{$N=2$}{N=2} sigma-models}\label{beta-sec}
Let us remind basic concepts of $\beta-$functions in pure metric sigma-models. The $\beta$-function $\beta_{ij}$ admits the covariant loop expansion
\begin{equation}\label{SM-beta-function-loop-expansion}
  \beta_{ij}=\beta_{ij}^{(1)}+\beta_{ij}^{(2)}+\beta_{ij}^{(3)}+\ldots\,,
\end{equation}
where the $L$-th loop order $\beta$-function coefficient $\beta_{ij}^{(L)}$ is a linear combination of tensors with given degree in the Planck constant $\hbar$. Namely,  since the metric $G_{ij}\sim\hbar^{-1}$ is proportional to the inverse of $\hbar$, it provides natural scaling properties of all tensors
\begin{equation}\label{hbar-counting}
   G_{ij}\sim\hbar^{-1}\implies G^{ij}\sim\hbar,\;\Gamma^{k}_{ij}\sim \hbar^0,\;\nabla_i\sim\hbar^0,\;
   R_{ijk}^{\quad l}\sim\hbar^{0},\;R_{ij}\sim \hbar^0\quad\text{etc}.\,,
\end{equation}  
Withing this scaling the $L-$th loop coefficient is a linear combination of tensors with
\begin{equation}
    \beta_{ij}^{(L)}\sim\hbar^{L-1}\,.
\end{equation}

For bosonic sigma-models first four coefficients $\beta_{ij}^{(L)}$ have been computed in the MS scheme  \cite{Ecker:1971xko,Friedan:1980jm,Graham:1987ep,Foakes:1987ij,Jack:1989vp}
\begin{equation}\label{SM-beta-function}
\begin{gathered}
    \beta_{ij}^{(1)}=R_{ij}\,,\quad
   \beta_{ij}^{(2)}=\frac{1}{2}R_{iklm}R_j^{\,\,klm}\,,\\
   \begin{multlined}
   \beta_{ij}^{(3)}=\frac{1}{8}\nabla_kR_{ilmn}\nabla^kR_j^{\,\,lmn}-\frac{1}{16}\nabla_iR_{klmn}\nabla_jR^{klmn}-\\
   -\frac{1}{2}R_{imnk}R_{jpq}^{\quad k}R^{mqnp}-\frac{3}{8}R_{iklj}R^{kmnp}R^{l}_{\,\,mnp}\,.
\end{multlined}
\end{gathered} 
\end{equation}
and 
\begin{equation}
    \beta_{ij}^{(4)}=T_{ij}+\zeta(3)\tilde{T}_{ij},
\end{equation}
where the explicit forms of the tensors $T_{ij}$ and $\tilde{T}_{ij}$ are given in \cite{Jack:1989vp}.

The $N=1$ supersymmetric $\beta$-function is also known in four loops and has much simpler form \cite{AlvarezGaume:1980dk,Alvarez-Gaume:1981exa,Grisaru:1986dk,Grisaru:1986px,Grisaru:1986wj}
\begin{equation}\label{SM-beta-function-SUSY}
   \beta_{ij}^{(1)}=R_{ij}\,,\quad
   \beta_{ij}^{(2)}=0\,,\quad \beta_{ij}^{(3)}=0\,\quad\text{and}\quad
   \beta_{ij}^{(4)}=\zeta(3)\tilde{T}_{ij}.
\end{equation}
We see that the second and the third loop contributions vanish, while the fourth loop coefficient $\beta_{ij}^{(4)}$ coincides exactly with the corresponding $\zeta(3)$ term in the bosonic $\beta-$function.

The $N=2$ supersymmetry requires that the target space metric is K\"ahler \cite{Zumino:1979et}. The $\beta$-function is exactly the same as \eqref{SM-beta-function-SUSY}, but specialized to the K\"ahler geometry. In the following, we will use Latin indexes  $(i,j,\dots)$ for real coordinates and Greek $(\mu,\nu,\bar{\mu},\bar{\nu},\dots)$ for complex ones. By definition of the K\"ahler geometry there is a choice of the complex structure such that the metric  has the form
\begin{equation}
    G_{\mu\bar{\nu}}=\partial_{\mu}\partial_{\bar{\nu}}K=\nabla_{\mu}\nabla_{\bar{\nu}}K,
\end{equation}
where $K$ is the K\"ahler potential. Then \cite{Grisaru:1986px,Grisaru:1986dk}
\begin{equation}\label{SUSY-4-loop}
    \beta_{\mu\bar{\nu}}^{(1)}=R_{\mu\bar{\nu}}\,,\quad
    \beta_{\mu\bar{\nu}}^{(4)}=\nabla_{\mu}\nabla_{\bar{\nu}}\Delta K\,,
\end{equation}
where\footnote{Note that, as pointed out in \cite{Grisaru:1986kw}, $\zeta(3)\tilde{T}_{\mu\bar{\nu}}$ differ from $\nabla_{\mu}\nabla_{\bar{\nu}}\Delta K$ by a diffeomorphism term.}
\begin{equation}\label{Delta-K}
  \Delta K=\frac{\zeta(3)}{24} R_{ijkl}R^{i}\,_{mn}\,^l\left(R^{jnmk}+R^{jkmn}\right).
\end{equation}
In terms of the K\"ahler potential the RG equation takes the form
\begin{equation}
    \dot{K}=\frac{1}{2}\log\det G-\Delta K+\mathcal{O}(\hbar^4)\,.
\end{equation}

The higher loop coefficients $\beta_{ij}^{(L)}$ for $L>1$ are regularization scheme dependent. In the covariant approach different schemes are related  by covariant metric redefinitions
\begin{equation}\label{metric-redefinitions}
    G_{ij}\rightarrow G_{ij}+\sum_{k=0}^{\infty}G_{ij}^{(k)}\,,\quad\text{where}\quad
    G_{ij}^{(k)}\sim\hbar^k\,.
\end{equation}
Using \eqref{metric-redefinitions} one can try to bring the $\beta$-function to some normal form. This form however is expected to be quite complicated, having non-trivial contributions at all loops. This is supported by the analysis performed at lower orders  \cite{Metsaev:1987zx} which shows that there are tensor structures in the $\beta$-function that are scheme independent.  In \cite{Alfimov:2021sir} it has been shown that the four-loop bosonic $\beta-$function for $D=2$ target space can be reduced to the form  \eqref{all-loop-beta-function-first-4-loops}. In particular, the $\zeta(3)$ term is one of scheme invariants that can not be exterminated by the replacement \eqref{metric-redefinitions}.

In \cite{Alfimov:2021sir} we noticed that for integrable sigma-models with $2D$ target space the $\zeta(3)$ term vanishes. One can argue that physically acceptable backgrounds should not contain transcendental terms $\zeta(n)$. This is confirmed by the exact form of the cigar metric \cite{Dijkgraaf:1991ba} and by the exact metric for the sausage model found in \cite{Hoare:2019ark,Alfimov:2021sir}.  Thus it is tempting to check this conjecture for higher dimensional backgrounds ($D>2$). Here however, one faces computational problem since the number of scheme parameters is very large\footnote{Here we count only the number of physically acceptable scheme parameters which correspond to decomposable tensors.  Namely, assume that the metric is decomposable
\begin{equation*}
ds^2=A_{ij}(\boldsymbol{X})dX^idX^j+B_{ij}(\boldsymbol{Y})dY^idY^j,
\end{equation*}
then $T_{ij}$ is called decomposable if it satisfies $T_{ij}=T_{ij}^A+T_{ij}^B$.
Then there is one parameter
\begin{equation*}
    G_{ij}^{(1)}=\xi_1R_{ij}
\end{equation*}
at the  order $\hbar$, $5$ parameters at the order $\hbar^2$
\begin{equation*}
G^{(2)}_{ij}=b_1R_{iklm}R_j\,^{klm}+b_2R_{ikjl}R^{kl}+b_3R_{ik}R_j^k+b_4\nabla^2 R_{ij}+b_5\nabla_i\nabla_jR
\end{equation*}
and more that $100$ parameters at the order $\hbar^3$!}. In order to facilitate this problem we consider $N=2$ supersymmetric version of this problem. 

General change of scheme which preserves $N=2$ supersymmetry corresponds to the following change of the K\"ahler potential\footnote{See \cite{Fulling:1992vm} for the basis for the scalars up to order $\hbar^4$.}
\begin{multline}\label{K-generic-scheme-change}
    K\rightarrow K+c_1\log\det G+ c_2R +\underbrace{c_3R^2+c_4\bigl(R_{ijkl}\bigr)^2+c_5\bigl(R_{ij}\bigr)^2+c_6\nabla^2 R}_{\text{$4$ scalars of order $\hbar^2$}}+\\+
    \underbrace{c_7 R^3+\dots+c_{23}\nabla^2\nabla^2R}_{\text{$17$ scalars of order $\hbar^3$ }}+
    \underbrace{c_8 R^4+\dots+c_{99}\nabla^2\nabla^2\nabla^2R}_{\text{$92$ scalars of order $\hbar^4$ }}+\dots,
\end{multline}
where the scalars of order $\hbar^L$ contribute to the change of the $\beta$-function starting from the $(L+2)$th loop order. Thus working withing the four loop approximation one has to consider only the first line in \eqref{K-generic-scheme-change}.

On the other side, since any $N=2$ sigma-model is a specialization of $N=1$ one, it looks natural to consider only universal schemes, which do not require any tensors other than the Riemanian metric in their construction. In the current case, one has to retain only the scalars $\Phi$ in \eqref{K-generic-scheme-change} whose second derivative $\nabla_{\mu}\nabla_{\bar{\nu}}\Phi$ correspond to  tensors $T_{ij}^{\Phi}$ in real coordinates \emph{without explicit use of the complex structure}
\begin{equation}\label{Kahler-property}
    2\nabla_{\mu}\nabla_{\bar{\nu}}\Phi dz^{\mu}d\bar{z}^{\bar{\nu}}=T_{ij}^{\scriptscriptstyle{\Phi}}dx^idx\,.
\end{equation}
where $T_{ij}^{\scriptscriptstyle{\Phi}}$ is a function of the curvature and its derivatives.  It can also be rephrased as follows: we are interested in those metric redefinitions  $g_{ij}\rightarrow g_{ij}+T_{ij}$ which have only mixed indexes when restricted to K\"ahler manifolds, and these mixed components have the potential form $T_{\mu\bar{\nu}}=\nabla_{\mu}\nabla_{\bar{\nu}}\Phi$. In that case the metric redefinition can be interpreted as a redefinition of the K\"ahler potential $K\rightarrow K+\Phi$. 
Similar question has been addressed in \cite{AlvarezGaume:1980dk,Alvarez-Gaume:1981exa,Pope:1986hg} while attempting to classify all possible counter-terms in $N=2$ sigma-models. Following \cite{AlvarezGaume:1980dk,Alvarez-Gaume:1981exa,Pope:1986hg} we will call such tensors \emph{K\"ahler tensors.}

First two terms in \eqref{K-generic-scheme-change} correspond to one and two loop counter-terms of order $1/\epsilon$ and $1/\epsilon^2$ respectively \cite{Grisaru:1986px,Grisaru:1986dk} and hence are expected to be K\"ahler. Indeed, one has
\begin{equation}\label{first-Kahler-identity}
T_{ij}^{\scriptscriptstyle{\log\det G}}=-2R_{ij},\quad
T_{ij}^{\scriptscriptstyle{R}}=\nabla^2 R_{ij}+2R_{ikjl}R^{kl}-2R_{ik}R_j^{k}.
\end{equation}
Moreover, in appendix \ref{Identities-proof} we show that 
\begin{multline}\label{second-Kahler-identity}
    T_{ij}^{\scriptscriptstyle{R_{ij}^2}}=
    2\nabla^kR_{il}\nabla_kR_{j}\,^l-2\nabla^kR_{il}\nabla^lR_{jk}+\nabla_{i}R_{kl}\nabla^kR_{j}\,^l+
    \nabla_{j}R_{kl}\nabla^kR_{i}\,^l+\\+2R^{kl}\nabla_k\nabla_lR_{ij}
    +4R_{kilj}R^{k}\,_mR^{lm}+2R_{klmi}R^k\,_jR^{lm}+2R_{klmj}R^k\,_iR^{lm},
\end{multline}
and
\begin{equation}\label{third-Kahler-identity}
    T_{ij}^{\scriptscriptstyle{\nabla^2 R}}=\nabla^2T_{ij}^{\scriptscriptstyle{R}}-\frac{1}{2}
    \left(R_{i}\,^kT_{kj}^{\scriptscriptstyle{R}}+R_{j}\,^kT_{ki}^{\scriptscriptstyle{R}}\right)+R_{i}\,^k\,_j\,^lT_{kl}^{\scriptscriptstyle{R}}.
\end{equation}
We were unable to find similar expressions for $R^2$ and $R_{ijkl}^2$ and we believe that the corresponding tensors depend explicitly on the complex structure\footnote{We note also that the $R^2$ term can be eliminated from the point of view that it  corresponds to undecomposable tensor.}.  Using this fact as an assumption, we drop the corresponding terms and consider the following change of scheme
\begin{equation}\label{K-specific-scheme-change}
    K\rightarrow K+c_1\log\det G+ c_2R +c_3\bigl(R_{ij}\bigr)^2+c_4\nabla^2 R,
\end{equation} 
which can be written as the following change of the metric
\begin{equation}\label{g-specific-scheme-change}
    g_{ij}\rightarrow g_{ij}+c_1T_{ij}^{\scriptscriptstyle{\log\det G}}+
    c_2T_{ij}^{\scriptscriptstyle{R}} +
    c_3T_{ij}^{\scriptscriptstyle{R_{ij}^2}}
    +c_4T_{ij}^{\scriptscriptstyle{\nabla^2 R}}.
\end{equation}
Taking
\begin{equation}\label{special-parameters}
    c_2=-\frac{c_1^2}{2},\quad c_3=-\frac{2c_1^3}{3}+\frac{5\zeta(3)}{48}\quad\text{and}\quad
    c_4=-\frac{c_1^3}{4}+\frac{\zeta(3)}{48},
\end{equation}
and using tedious, but straightforward calculation, one shows that in the new scheme the K\"ahler potential satisfies  
\begin{equation}\label{new-K}
    \dot{K}=\frac{1}{2}\log\det G-\Delta\tilde{K}+\mathcal{O}(\hbar^4)\,,
\end{equation}
with\footnote{First equality in \eqref{special-parameters} provides that the $\beta$-function does not get corrected in the second and third loop orders.}
\begin{multline}\label{delta-K-tilde}
    \Delta\tilde{K} =\frac{\zeta(3)}{24}\left( R_{ijkl}R^{i}\,_{mn}\,^k\left(R^{jnml}+R^{jlmn}\right)-6R^{i j}R^{k l}R_{i k j l}
    +\right.\\
     \left.+R^{ij}R_{ik}R^k\,_j-3R^{i j} \nabla^{2} R_{i j}+\frac{3}{4} \nabla^{2} R_{i j}^2\right)\,.
\end{multline}
We shall call the corresponding choice of parameters in \eqref{g-specific-scheme-change} the integrable scheme.

The explicit form of the scalar \eqref{delta-K-tilde} is not very illuminating. Nevertheless, one can show that for $D=2$ it reduces to the left hand side of \eqref{R3-relation} up to a factor. Our main result states that $\Delta\tilde{K}$ given by \eqref{delta-K-tilde} is a constant for integrable K\"ahler sigma-models such as complete $T-$dual of $\eta-$deformed $\mathbb{CP}(n)$ model and thus does not give corrections to the metric $\beta-$function. 
%%%%%%%%%%%%%%%%%%%%%%%%%%%%%%%%%%%%%%%%%%%%%%%%%%%%%%%%%%%%%%%%%%%%%%%%%%%%%%%%%%%%%%%%%%%%%%%%%%%%%%%%%%%%%%%%%%%%%%%%%%%%%%%%%%%%%%%%%%%%%%%%%%%%%%%%%%%%%%%%%%%%%%%%%%%%%%%%%%%%%%%%%%%%%%%%%%%%%%%%%%%%%%%%%%%%%%%%%%%%%%%%%%%%%%%%%%%%%%%%%%%%%%%%%%%%%%%%%%%%%%%%%%%
\section{Metrics of  complete \texorpdfstring{$T-$}{T}dual to \texorpdfstring{$\eta-$}{eta-}deformed \texorpdfstring{$\mathbb{CP}(n-1)$}{CP(n-1)} model and the corresponding integrable sigma-model}\label{Tdual-CPn} 
The bosonic action of the $\eta-$deformed $G/H$ sigma-model has the form  \cite{Klimcik:2008eq,Delduc:2013fga}
\begin{equation}\label{Coset-action-deformed}
\mathcal{S}=\frac{\kappa}{4\pi \hbar}\int\textrm{Tr}\left(
\left(\mathbf{g}\partial_{+}\mathbf{g}^{-1}\right)^{(\textrm{c})}\,\frac{1}{1-i\kappa\mathcal{R}_{\mathbf{g}}\circ\mathrm{P}_{\textrm{c}}}\,
\left(\mathbf{g}\partial_{-}\mathbf{g}^{-1}\right)^{(\textrm{c})}\right) d^{2}x\,,
\end{equation}
where $\mathcal{R}$ is the Drinfel’d-Jimbo solution to the modified YB equation $\mathbf{g}\in G$, $\mathcal{R}_{\mathbf{g}}=\textrm{Ad}\, \mathbf{g}\circ\mathcal{R}\circ\textrm{Ad}\,\mathbf{g}^{-1}$ and $\mathrm{P}_{\textrm{c}}$ is the projector on the coset space. In the case when $G/H$ is a symmetric space,  the model \eqref{Coset-action-deformed} is known to be one loop renormalizable with only one running coupling \cite{Hoare:2021dix} 
\begin{equation}\label{kappa-equation}
    \dot{\kappa}=\frac{c_{2}(G)\hbar(\kappa^2-1)}{4},
\end{equation}
where $c_{2}(G)$ is the dual Coxeter number of $G$.

We consider the theory \eqref{Coset-action-deformed} for $G=SU(n)$ and $H=U(n-1)$, which corresponds to deformation of $\mathbb{CP}(n-1)$ sigma model. In this case $C_2(G)=n$ and the solution of  \eqref{kappa-equation} has the form
\begin{equation}\label{kappa-explicit}
    \kappa=\frac{1-e^{\frac{n\hbar}{2}t}}{1+e^{\frac{n\hbar}{2}t}}
\end{equation}
The theory contains both the metric and the $B-$field (except for the case of $n=1$). However, as it was observed in \cite{Litvinov:2019rlv,Bykov:2020llx}, the $T-$duality along all the isometric directions (complete $T-$dual) eliminates the $B-$field completely and makes the metric rather simple. Moreover in \cite{Bykov:2020llx} it has been shown that the corresponding geometry is K\"ahler. 

The corresponding metrics from \cite{Litvinov:2019rlv} and from \cite{Bykov:2020llx} are written in different coordinates. The metric  found in \cite{Litvinov:2019rlv} has a nice form of the flat metric perturbed by the "graviton-like" exponential terms (here for simplicity  $\hbar=1$)
\begin{equation}\label{metric-1}
    ds^2=|d\boldsymbol{z}|^2+\frac{2}{e^{nt}-1}\sum_{k=1}^nf_k(\boldsymbol{x})(\boldsymbol{h}_k\cdot d\boldsymbol{z})^2, \quad \boldsymbol{z}=\boldsymbol{x} + i \boldsymbol{y},\quad
    \boldsymbol{z}=\big(z_1,\dots,z_{n-1}\big),
\end{equation}
where  
\begin{equation}
    f_k(\boldsymbol{x})=\sum_{l=1}^n e^{l t} e^{((\boldsymbol{\alpha}_k + \dots+\boldsymbol{\alpha}_{k+l-1})\cdot \boldsymbol{x})}\,.
\end{equation}
Here $\boldsymbol{\alpha}_k,\, k=1,\dots,n-1$ are the simple roots of $\mathfrak{sl}(n)$ 
\begin{equation}\label{roots}
  \boldsymbol{\alpha}_n=-\sum_{k=1}^{n-1}\boldsymbol{\alpha}_k,\quad
  \boldsymbol{\alpha}_{n+k}\equiv\boldsymbol{\alpha}_k\,,
\end{equation}
and $\boldsymbol{h}_k$ are the weights  
\begin{equation}\label{weights}
    \boldsymbol{h}_1=\boldsymbol{\omega}_1,\quad
    \boldsymbol{h}_2=\boldsymbol{\omega}_1-\boldsymbol{\alpha}_1,\quad
    \boldsymbol{h}_2=\boldsymbol{\omega}_1-\boldsymbol{\alpha}_1-\boldsymbol{\alpha}_2,\quad\dots,\quad (\boldsymbol{\omega}_i,\boldsymbol{\alpha}_j)=\delta_{ij}.
\end{equation}
This metric was constructed in \cite{Litvinov:2019rlv} as a conjectured dual to the bosonic $\mathfrak{gl}(n|n)$ Toda field theory. Among other things it was constructed so that it solves Ricci flow equation \eqref{RG-equation}.

The metric from \cite{Bykov:2020llx} is K\"ahler and provided by the potential
\begin{equation}
    K(\boldsymbol{\zeta},\bar{\boldsymbol{\zeta}})= \sum_{k=2}^{n-1} i (\boldsymbol{\zeta}_k \bar{\boldsymbol{\zeta}}_{k-1}-\boldsymbol{\zeta}_{k-1} \bar{\boldsymbol{\zeta}}_k) + 2 \sum_{k=1}^n P(\boldsymbol{t}_k - \boldsymbol{t}_{k-1} + 2 \tau), \quad \boldsymbol{\zeta} = \boldsymbol{\xi}+i\boldsymbol{\psi},
\end{equation}
where $P(\rho) \equiv i \left(\text{Li}_2 (e^{i \rho})\,+\,\frac{\rho(2 \pi - \rho)}{4}\right)$,  $\boldsymbol{t}_k = \boldsymbol{\zeta}_k+\bar{\boldsymbol{\zeta}}_k$ for $k = 1 \dots n-1$ and $\boldsymbol{t}_0 = - \boldsymbol{t}_n$. It is also convenient to make the following change, in order to get rid of $\boldsymbol{t}_n$:
\begin{equation}
    \boldsymbol{\xi}_k \rightarrow \boldsymbol{\xi}_k + \boldsymbol{t}_n \left(\frac{k}{n}-\frac{1}{2}\right), \quad \tau \rightarrow \tau - \frac{\boldsymbol{t}_n}{n}.
\end{equation}
We have found the following simple change of variables, which relates two metrics
\begin{equation}
    e^{(\boldsymbol{\alpha}_k\cdot\boldsymbol{x})}=\frac{e^{t}e^{2 i \boldsymbol{\xi}_k}-e^{2 i \boldsymbol{\xi}_{k-1}}}{e^{t}e^{2 i \boldsymbol{\xi}_{k-1}}-e^{2 i \boldsymbol{\xi}_{k-2}}}, \quad (\boldsymbol{h}_k \cdot\boldsymbol{z}) = i \boldsymbol{\zeta}_{k-1}  - i \boldsymbol{\zeta}_{k-2}, \quad k = 1 \dots n-1, 
\end{equation}
where 
\begin{equation}
 \boldsymbol{\xi}_{n+k} \equiv \boldsymbol{\xi}_{k},\quad
 \boldsymbol{\zeta}_{n+k} \equiv \boldsymbol{\zeta}_{k},\quad 
 \boldsymbol{\zeta}_{0} = \boldsymbol{\xi}_{0} = 0,\quad \tau = -\frac{i t}{2}   \quad
 \sum_{k=1}^n\boldsymbol{\alpha}_k=0.
\end{equation}

Using explicit calculation with the help of Mathematica package xAct \cite{Nutma:2013zea}, one finds  that $\Delta\tilde{K}$ given by \eqref{delta-K-tilde} is indeed a constant for the  background \eqref{metric-1}. Namely, it can be shown that (compare to \eqref{R3-relation})
\begin{multline}\label{R3-relation-higher-D}
    \left(R_{ijkl}R^{i}\,_{rs}\,^k\big(R^{jsrl}+R^{jlrs}\big)-6R^{i j}R^{k l}R_{i k j l}+R^{ij}R_{ik}R^k\,_j-3R^{i j} \nabla^{2} R_{i j}+\frac{3}{4} \nabla^{2} R_{i j}^2\right)=\\=
    -\frac{3}{16}n^2(n-1)\hbar^3\big(\kappa-\kappa^{-1}\big)^2\big(\kappa+\kappa^{-1}\big),
\end{multline}
where $\kappa$ is the parameter of the metric \eqref{kappa-explicit}. This formula generalizes \eqref{R3-relation} for $D>2$.

Now we present some arguments that the supersymmetric sigma-model with the metric  \eqref{metric-1} is one-loop exact.  First, we comment on the bosonic model corresponding to \eqref{metric-1}. While being classically integrable, it is anomalous at the quantum level similarly to the undeformed $\mathbb{CP}(n-1)$ model \cite{Abdalla:1980jt}. However, as it was shown in \cite{Litvinov:2019rlv,Fateev:2019xuq},  quantum integrability is restored if one adds additional bosonic field that decouples in the classical limit. The corresponding metric that includes this  degree of freedom admits the UV expansion \cite{Litvinov:2019rlv}
\begin{equation}\label{UV-expansion-bosonic-metric}
    ds^2=\sum_{i=1}^{n-1}dx_i^2+\sum_{i=1}^{n}dy_i^2+e^{t}
    \sum_{k=1}^n\Big|\big(\boldsymbol{h}_k\cdot d\boldsymbol{x}\big)
    -i\sqrt{1+b^{-2}}\big(\boldsymbol{\mathfrak{h}}_k\cdot d\boldsymbol{y}\big)\Big|^2
    e^{\frac{(\boldsymbol{\alpha}_k\cdot \boldsymbol{x})}{b}}+O(e^{2t}),
\end{equation}
where $\boldsymbol{\alpha}_k$ and $\boldsymbol{h}_k$ are given by \eqref{roots}-\eqref{weights} and
\begin{equation}
    (\boldsymbol{\mathfrak{h}}_1,\dots,\boldsymbol{\mathfrak{h}}_n)\in\mathbb{R}^n:\qquad
    \big(\boldsymbol{\mathfrak{h}}_i\cdot\boldsymbol{\mathfrak{h}}_j\big)=\delta_{ij}-\frac{1}{1+b^{-2}}\frac{1}{n}.
\end{equation}
It is important that the leading coefficients in \eqref{UV-expansion-bosonic-metric} are exact in the Planck constant $\hbar\overset{\text{def}}{=}b^{-2}$. We note that in the semiclassical limit the basis $\boldsymbol{\mathfrak{h}}_k$ is degenerate 
\begin{equation}
    \big(\boldsymbol{\mathfrak{h}}_i\cdot\boldsymbol{\mathfrak{h}}_j\big)=\delta_{ij}-\frac{b^2}{1+b^2}\frac{1}{n}\,\longrightarrow\delta_{ij}-\frac{1}{n}\implies \Big(\sum_{k=1}^n\boldsymbol{\mathfrak{h}}_k\cdot\boldsymbol{\mathfrak{h}}_i\Big)\rightarrow0.
\end{equation}
and hence after rescaling $\boldsymbol{x}\rightarrow b\boldsymbol{x}$, $\boldsymbol{y}\rightarrow b\boldsymbol{y}$ one degree of freedom decouples and one arrives to \eqref{metric-1} after simple change of variables. The form of the leading coefficient in \eqref{UV-expansion-bosonic-metric} is constrained by integrability. The corresponding integrable system  is defined by the  screening operators for bosonic $\mathfrak{gl}(n|n)$ Toda QFT \cite{Litvinov:2019rlv,Fateev:2019xuq}
\begin{equation}\label{screenings}
    \mathcal{S}_{2r-1}=\oint e^{b(\boldsymbol{h}_k\cdot\boldsymbol{x}(z))-i\sqrt{1+b^2}(\boldsymbol{\mathfrak{h}}_k\cdot\boldsymbol{y}(z))}dz,\quad
    \mathcal{S}_{2r}=\oint e^{i\sqrt{1+b^2}(\boldsymbol{\mathfrak{h}}_k\cdot\boldsymbol{y}(z))-b(\boldsymbol{h}_{k+1}\cdot\boldsymbol{x}(z))}dz\,,
\end{equation}
with $r=1,\dots,n$.

Contrary, the corresponding supersymmetric model\footnote{
The supersymmetric action is defined in terms of $N=1$ super field as ususal
\begin{equation*}
S=\frac{1}{4\pi}\int G_{ij}(\mathbb{X})\mathcal{D} \mathbb{X}^{i}\bar{\mathcal{D}}\mathbb{X}^{j}\,d^{2}\boldsymbol{x}\,d^2\theta.
\end{equation*}
where
\begin{equation*}
 \mathbb{X}^i(z,\bar{z},\theta,\bar{\theta})\overset{\text{def}}{=}
  X^i(z,\bar{z})+\theta\psi^i(z,\bar{z})+\bar{\theta}\bar{\psi}^{i}(z,\bar{z})+
  \theta\bar{\theta}F^i(z,\bar{z}),\quad
  \mathcal{D}\overset{\text{def}}{=}\frac{\partial}{\partial\theta}-\theta\partial,\quad
  \bar{\mathcal{D}}\overset{\text{def}}{=}\frac{\partial}{\partial\bar{\theta}}-\bar{\theta}\bar{\partial}\,.
\end{equation*}
}
is integrable as it is, without auxiliary degrees of freedom. The metric has the UV expansion
\begin{equation}\label{UV-expansion-SUSY-metric}
    ds^2=\sum_{i=1}^{n-1}dx_i^2+\sum_{i=1}^{n-1}dy_i^2+e^{t}
    \sum_{k=1}^n\Big|\big(\boldsymbol{h}_k\cdot d\boldsymbol{x}\big)
    -i\big(\boldsymbol{h}_k\cdot d\boldsymbol{y}\big)\Big|^2
    e^{\frac{(\boldsymbol{\alpha}_k\cdot \boldsymbol{x})}{b}}+O(e^{2t}),
\end{equation}
which is controlled by the set of screening fields for supersymmetric $\mathfrak{gl}(n|n)$ Toda theory \cite{Olshanetsky:1982sb,Evans:1990qq,Ito:1991wb} 
\begin{equation}\label{screenings-SUSY}
    \tilde{\mathcal{S}}_{2r-1}=\oint e^{b(\boldsymbol{h}_k\cdot\boldsymbol{x}(z,\theta))-ib(\boldsymbol{h}_k\cdot\boldsymbol{y}(z,\theta))}dzd\theta,\quad
    \mathcal{S}_{2r}=\oint e^{ib(\boldsymbol{h}_k\cdot\boldsymbol{y}(z,\theta))-b(\boldsymbol{h}_{k+1}\cdot\boldsymbol{x}(z,\theta))}dz\theta\,.
\end{equation}
We note that the set of screening fields \eqref{screenings-SUSY} defines integrable perturbation of Kazama-Suzuki $N=2$ coset model for $SU(n)/U(n-1)$ \cite{Kazama:1988qp}.

Comparing the systems \eqref{screenings} and \eqref{screenings-SUSY} one finds that the later system coincides with the limit $\hbar=b^{-2}\rightarrow0$ of the former. Moreover the form of \eqref{UV-expansion-SUSY-metric} and \eqref{screenings-SUSY} suggests that the corresponding sigma-model is one-loop exact. On the other side \eqref{UV-expansion-bosonic-metric} and \eqref{screenings} show that the bosonic action will receive corrections at every loop order. Of course the conjecture that the supersymmetric sigma-model governed by the UV expansion \eqref{UV-expansion-SUSY-metric} is one-loop exact has to be confirmed by explicit calculations. In this paper we have shown, that there exits a scheme in which the fourth loop coefficient of the $\beta-$function vanishes.
%%%%%%%%%%%%%%%%%%%%%%%%%%%%%%%%%%%%%%%%%%%%%%%%%%%%%%%%%%%%%%%%%%%%%%%%%%%%%%%%%%%%%%%%%%%%%%%%%%%%%%%%%%%%%%%%%%%%%%%%%%%%%%%%%%%%%%%%%%%%%%%%%%%%%%%%%%%%%%%%%%%%%%%%%%%%%%%%%%
\section{Conclusion}\label{conclusion}
In these notes we studied scheme dependence of $N=2$ supersymmetric pure metric sigma-models. We have shown that for integrable backgrounds such as complete $T-$dual of $\eta-$deformed $\mathbb{CP}(n-1)$ sigma-models there is a regularization scheme in which the fourth-loop contribution to the $\beta-$function vanishes, thus making it one loop exact to this order.
There are several immediate open questions, which we mention 
\paragraph{$\lambda$-model.}
In our previous paper \cite{Alfimov:2021sir} it has been also shown that \eqref{R3-relation} holds also for $\lambda$-deformed model \cite{Sfetsos:2013wia,Hollowood:2014rla} for $SU(2)/U(1)$.  It is natural to expect that higher dimensional generalization of  \eqref{R3-relation}, namely \eqref{R3-relation-higher-D}, holds for $SU(n)/U(n-1)$ $\lambda$-model. We have checked explicitly that this is the case for $n=3$. Let us briefly comment on this calculation. 

It is well known that $\eta$ and $\lambda$ deformations of $G/H$ sigma-model are related by Poisson-Lie duality with respect to group $G$ and certain analytic continuation \cite{Hoare:2017ukq}. There is another relation between two models noticed in \cite{Hoare:2015gda} for $SO(n+1)/SO(n)$. Namely,  taking certain infinite limits of the $\lambda-$model one recovers the complete abelian $T-$dual of the $\eta-$model. It is not automatically clear  that for any coset $G/H$ it is always possible to find limits that provide similar relation. Thus in general $\lambda-$model provides independent check of our conjecture. 

The action of the $G/H$ $\lambda-$model has the form (here $g\in G$)
\begin{equation}
   S=\int\textrm{Tr}\Bigl(-\frac{1}{2}(g^{-1}\partial g)(g^{-1}\bar{\partial}g)+J\left(\textrm{Ad}_{g}-1+\lambda\mathbb{P}\right)^{-1}\bar{J}\Bigr)d^{2}\xi+S_{WZ},
\end{equation}
where $\mathbb{P}$ is the projection on the coset space, in our case $SU(3)/U(2)$, and
\begin{equation*}
   J=g^{-1}\cdot\partial g\,, \quad \bar{J}=\bar{\partial}g\cdot g^{-1}\,.
\end{equation*}
We have taken the following parametrization of the group element
\begin{equation}
    g=e^{i\alpha t_1}e^{i\beta t_6}e^{i\gamma t_3}e^{i\delta t_8},
\end{equation}
where $t_k$ are the standard Gell-Mann matrices, one finds that the model has a non-vanishing, but  constant $B-$field
\begin{equation}
    B=-\frac{1}{6}d\gamma\wedge d\delta.
\end{equation}
The corresponding metric is very complicated to be presented here. However,  we have checked that this metric is K\"ahler and  $\Delta\tilde{K}=\text{const}$. It would be interesting to repeat this calculation for general $n$.
%%%%%%%%%%%%%%%%%%%%%%%%%%%%%%%%%%%%%%%%%%%%%%%%%%%%%%%%%%%%%%%%%%%%%%%%%%%%%%%%%%%%%%%%%%
\paragraph{Fifth loop computations}
It will be interesting to extend our results to fifth loop. The corresponding contribution to the $\beta$-function has been computed in \cite{Grisaru:1986wj}
\begin{equation}
    \beta_{\mu\bar{\nu}}^{(5)}=\frac{\zeta(4)}{48}\nabla_{\mu}\nabla_{\bar{\mu}}\nabla^2
    R_{ijkl}R^{i}\,_{mn}\,^l\left(R^{jnmk}+R^{jkmn}\right).
\end{equation}
In order to study the scheme dependence at this order one has to consider more general redefinitions of the K\"ahler potential which include scalars of order $\hbar^3$ in \eqref{K-generic-scheme-change}. For generic Riemanian metric there are $17$ such scalars \cite{Fulling:1992vm}\footnote{Note that for a given $D$ there might be more relations (see e.g. \cite{Xu:1987pz} for $D=4$)}. Note that in the case of K\"ahler geometry there is the relation (see appendix \ref{Identities-proof} for details)
\begin{equation}\label{Kahler-scalar-relation-1}
R^{ij}\nabla_{i}\nabla_jR=R^{ij}\nabla^2R_{ij}+2R^{ij}R_{ikjl}R^{kl}-2R^{ij}R_{ik}R^k\,_j
\end{equation}
which reduces the number of independent scalars to $16$. 

In order to deal with universal scheme redefinitions, one has to consider only the scalars whose mixed derivatives correspond to K\"ahler tensors. It is expected that one reduces the fifth loop beta function to a form similar to \eqref{delta-K-tilde}, which vanishes for integrable backgrounds. Apart from trivial contributions of the form $\nabla^2\Delta\tilde{K}$, that vanish due to \eqref{R3-relation-higher-D}, we expect high order relations. In the case of $D=2$ we have found at least one relation at every order in $\hbar$ starting from $\hbar^3$. In paeticular,  at the order $\hbar^4$ one has
\begin{multline}\label{R4-relation}
    R^4+3R^2\nabla^2R-\frac{1}{6}R\nabla^2R^2-\frac{1}{9}\nabla^2R^3+\frac{2}{3}
    \left(\big(\nabla^2R\big)^2+R\nabla^2\nabla^2R\right)-\frac{1}{3}\nabla^2\left(R\nabla^2R\right)=\\=
    \frac{\hbar^4}{3}\big(\kappa-\kappa^{-1}\big)^2\big(3\kappa^2+3\kappa^{-2}+2\big)\,,
\end{multline}
valid for integrable $\eta$ and $\lambda$-deformed $SU(2)/U(1)$ models.
%%%%%%%%%%%%%%%%%%%%%%%%%%%%%%%%%%%%%%%%%%%%%%%%%%%%%%%%%%%%%%%%%%%%%%%%%%%%%%%%%%%%%%%%%%
%%%%%%%%%%%%%%%%%%%%%%%%%%%%%%%%%%%%%%%%%%%%%%%%%%%%%%%%%%%%%%%%%%%%%%%%%%%%%%%%%%%%%%%%%%
\section*{Acknowledgments}
We acknowledge discussions with Alexey Rosly. A.L. has been supported by the Russian Science Foundation under the grant 22-22-00991 and by Basis foundation. A.L. is also grateful to Abubakir Koshek for collaboration at the early stages of this project.
%%%%%%%%%%%%%%%%%%%%%%%%%%%%%%%%%%%%%%%%%%%%%%%%%%%%%%%%%%%%%%%%%%%%%%%%%%%%%%%%%%%%%%%%%%%%%%%%%%%%%%%%%%%%%%%%%%%%%%%%%%%%%%%%%%%%%%%%%%%%%%%%%%%%%%%%%%%%%%%%%%%%%%%%%%%%%%%%%%%%%%%%%%%%%%%%%%%%%%%%%%%%%%%%%%%%%%%%%%%%%%%%%%%%%%%%%%%%%%%%%%%%%%%%%%%%%%%%%%%%%%%%%%%%%%%%%%%%%%%%%%%%%%%%
\appendix
\section{K\"ahler tensors}\label{Identities-proof}
K\"ahler metric satisfies special properties. In particular the only non–vanishing components
of the curvature are the mixed ones
\begin{equation}
    R_{\alpha\bar{\beta}\gamma\bar{\delta}}\quad\text{etc}\,,
\end{equation}
which leads to the following identities  
\begin{equation}\label{Identity-1}
    \nabla_{\mu}R_{\alpha\bar{\alpha}}=\nabla_{\alpha}R_{\mu\bar{\alpha}},\quad
    \nabla_{\bar{\mu}}R_{\alpha\bar{\alpha}}=\nabla_{\bar{\alpha}}R_{\alpha\bar{\mu}}\,.
\end{equation}

Consider a scalar $\Phi$ that satisfies the K\"ahler property \eqref{Kahler-property}. We have
\begin{multline}
\nabla_{\mu}\nabla_{\bar{\nu}}\nabla^2\Phi=2g^{\alpha\bar{\alpha}}\nabla_{\mu}\nabla_{\bar{\nu}}\nabla_{\alpha}\nabla_{\bar{\alpha}}\Phi=2g^{\alpha\bar{\alpha}}\nabla_{\mu}\nabla_{\bar{\alpha}}\nabla_{\alpha}\nabla_{\bar{\nu}}\Phi=\\=
2g^{\alpha\bar{\alpha}}\left(\nabla_{\bar{\alpha}}\nabla_{\alpha}\nabla_{\mu}\nabla_{\bar{\nu}}\Phi+R_{\mu\bar{\alpha}\alpha}\,^p\nabla_{p}\nabla_{\bar{\nu}}\Phi
+R_{\mu\bar{\alpha}\bar{\nu}}\,^p\nabla_{\alpha}\nabla_{p}\Phi\right)=\\=
2g^{\alpha\bar{\alpha}}\nabla_{\bar{\alpha}}
\nabla_{\alpha}\nabla_{\mu}\nabla_{\bar{\nu}}\Phi-R_{\mu}\,^p\nabla_p\nabla_{\bar{\nu}}\Phi+R_{\bar{\nu}k\mu l}\nabla^k\nabla^l\Phi,
\end{multline}
and at the same time exchanging the role of indexes $\mu$ and $\bar{\nu}$ we have
\begin{equation}
\nabla_{\mu}\nabla_{\bar{\nu}}\nabla^2\Phi=
2g^{\alpha\bar{\alpha}}\nabla_{\alpha}
\nabla_{\bar{\alpha}}\nabla_{\mu}\nabla_{\bar{\nu}}\Phi-R_{\bar{\nu}}\,^p\nabla_p\nabla_{\mu}\Phi+R_{\mu k\bar{\nu} l}\nabla^k\nabla^l\Phi.
\end{equation}
Taking symmetric combination, we obtain
\begin{equation}
    \nabla_{\mu}\nabla_{\bar{\nu}}\nabla^2\Phi=\nabla^2\nabla_{\mu}\nabla_{\bar{\nu}}\Phi-\frac{1}{2}\left(R_{\mu}\,^p\nabla_p\nabla_{\bar{\nu}}\Phi+
    R_{\bar{\nu}}\,^p\nabla_p\nabla_{\mu}\Phi\right)+R_{\mu k\bar{\nu} l}\nabla^{k}\nabla^{l}\Phi.
\end{equation}
Finally, using \eqref{Kahler-property} we find the relation
\begin{equation}
    T_{ij}^{\scriptscriptstyle{\nabla^2 \Phi}}=\nabla^2T_{ij}^{\scriptscriptstyle{\Phi}}-\frac{1}{2}
    \left(R_{i}\,^kT_{kj}^{\scriptscriptstyle{\Phi}}+R_{j}\,^kT_{ki}^{\scriptscriptstyle{\Phi}}\right)+R_{i}\,^k\,_j\,^lT_{kl}^{\scriptscriptstyle{\Phi}}
\end{equation}
Taking for $\Phi=\log\det g$ and for $\Phi=R$ we obtain exactly \eqref{first-Kahler-identity} and \eqref{third-Kahler-identity}. We note that contracting \eqref{first-Kahler-identity} with $R^{ij}$, one finds \eqref{Kahler-scalar-relation-1}.

Now we prove K\"ahler identity for $R_{ij}^2$. Consider 
\begin{equation}\label{I1-I2}
    \nabla_{\mu}\nabla_{\bar{\nu}}R_{ij}^2=2\nabla_{\mu}R_{ij}
    \nabla_{\bar{\nu}}R^{ij}
    +2\left(\nabla_{\mu}\nabla_{\bar{\nu}}R_{ij}\right)R^{ij}=I_1+I_2.
\end{equation}
The first tensor in \eqref{I1-2} can be transformed either as
\begin{multline}\label{I1-2}
    I_1=4\nabla_{\mu}R_{\alpha\bar{\alpha}}
    \nabla_{\bar{\nu}}R^{\alpha\bar{\alpha}}
    \overset{\eqref{Identity-1}}{=}
    4\nabla_{\alpha}R_{\mu\bar{\alpha}}
    \nabla^{\alpha}R_{\bar{\nu}}\,^{\bar{\alpha}}=
    4\nabla_{k}R_{\mu l}\nabla^{k}R_{\bar{\nu}}\,^{l}-
    4\nabla_{\bar{\beta}}R_{\mu\bar{\alpha}}\nabla^{\bar{\beta}}R_{\bar{\nu}}\,^{\bar{\alpha}}
    \overset{\eqref{Identity-1}}{=}\\\overset{\eqref{Identity-1}}{=}
    4\nabla_{k}R_{\mu l}\nabla^{k}R_{\bar{\nu}}\,^{l}-
    4\nabla_{\bar{\alpha}}R_{\mu\bar{\beta}}\nabla^{\bar{\beta}}R_{\bar{\nu}}\,^{\bar{\alpha}}
    =4\nabla_{k}R_{\mu l}\nabla^{k}R_{\bar{\nu}}\,^{l}-4\nabla_{k}R_{\mu l}\nabla^{l}R_{\bar{\nu}}\,^{k},
\end{multline}
or as
\begin{multline}\label{I1-1}
    I_1=4\nabla_{\mu}R_{\alpha\bar{\alpha}}
    \nabla_{\bar{\nu}}R^{\alpha\bar{\alpha}}
    \overset{\eqref{Identity-1}}{=}
    2\nabla_{\alpha}R_{\mu\bar{\alpha}}
    \nabla_{\bar{\nu}}R^{\alpha\bar{\alpha}}+2\nabla_{\mu}R_{\alpha\bar{\alpha}}\nabla^{\alpha}R_{\bar{\nu}}\,^{\bar{\alpha}}=\\=
    2\nabla_{\mu}R_{kl}\nabla^kR_{\bar{\nu}}\,^l+
    2\nabla_{\bar{\nu}}R_{kl}\nabla^kR_{\mu}\,^l,
\end{multline}
Taking symmetric combination of \eqref{I1-2} and \eqref{I1-1} one finds the first line in \eqref{second-Kahler-identity} written in real coordinates. On the other hand, using  the identity \eqref{Identity-1}, one can easily show that this tensor has only mixed components. 
The second term $I_2$ in \eqref{I1-2} we can represent ways
\begin{multline}\label{I2-1}
    I_2=4\left(\nabla_{\mu}\nabla_{\bar{\nu}}R_{\alpha\bar{\alpha}}\right)R^{\alpha\bar{\alpha}}
    \overset{\eqref{Identity-1}}{=}
    4\left(\nabla_{\mu}\nabla_{\bar{\alpha}}R_{\alpha\bar{\nu}}\right)R^{\alpha\bar{\alpha}}=\\=
    4\left(\nabla_{\bar{\alpha}}\nabla_{\alpha}R_{\mu\bar{\nu}}+
    R_{\bar{\nu}\alpha\mu}\,^pR_{p\bar{\alpha}}+
    R_{\bar{\nu}\alpha\bar{\alpha}}\,^pR_{\mu p}\right)R^{\alpha\bar{\alpha}},
\end{multline}
and
\begin{multline}\label{I2-2}
    I_2=4\left(\nabla_{\bar{\nu}\nabla_{\mu}}R_{\alpha\bar{\alpha}}\right)R^{\alpha\bar{\alpha}}
    \overset{\eqref{Identity-1}}{=}
    4\left(\nabla_{\bar{\nu}}\nabla_{\alpha}R_{\mu\bar{\alpha}}\right)R^{\alpha\bar{\alpha}}=\\=
    4\left(\nabla_{\alpha}\nabla_{\bar{\alpha}}R_{\mu\bar{\nu}}+
    R_{\mu\bar{\alpha}\alpha}\,^pR_{p\bar{\nu}}+
    R_{\mu\bar{\alpha}\bar{\nu}}\,^pR_{\alpha p}\right)R^{\alpha\bar{\alpha}}.
\end{multline}
Taking symmetric combination of \eqref{I2-1} and \eqref{I2-2} we find the second line in \eqref{second-Kahler-identity} which also has only mixed components. 
\bibliographystyle{JHEP}
\bibliography{MyBib}

\end{document}